\documentclass{ws-procs9x6}

\newcommand{\bbar}[1]{{\,\overline{\!#1}{}}}

\newcommand{\orderone}{$\mathcal{O}(1)$\ }

\newcommand{\beq}{\begin{equation}}
\newcommand{\eeq}{\end{equation}}
\newcommand{\beqn}{\begin{eqnarray}}
\newcommand{\eeqn}{\end{eqnarray}}

\begin{document}

\title{Proton Decay and the Planck Scale\footnote{\uppercase{T}alk
presented at \emph{\uppercase{PASCOS} '04}, held at
\uppercase{N}ortheastern \uppercase{U}niversity, \uppercase{B}oston,
\uppercase{MA}, \uppercase{A}ugust 16-22,
2004. \uppercase{T}o appear in the \uppercase{P}roceedings.}}

\author{DANIEL T. LARSON}

\address{Theoretical Physics Group, 
Lawrence Berkeley National Laboratory, 
University of California, Berkeley, CA 94720, USA}
\address{Department of Physics, University of California,
Berkeley, CA 94720, USA}

\maketitle

\abstracts{
Even without grand unification, proton decay can be a powerful probe
of physics at the highest energy scales. Supersymmetric theories with
conserved $R$-parity contain Planck-suppressed dimension 5 operators
that give important contributions to nucleon decay. These operators
are likely controlled by flavor physics, which means current and near
future proton decay experiments might yield clues about the fermion
mass spectrum. I present a thorough analysis of nucleon partial
lifetimes in supersymmetric one-flavon Froggatt-Nielsen models with a
single $U(1)_X$ family symmetry which is responsible for the fermionic
mass spectrum as well as forbidding $R$-parity violating interactions.
Many of the models naturally lead to nucleon decay near present limits
without any reference to grand unification.}

\section{Two Myths}
\label{sec:myths}

It is often loosely stated that the observation of proton decay
implies the existence of a grand unified theory (GUT). However, it is
well known that generic supersymmetric (SUSY) theories possess
nonrenormalizable operators that violate baryon- and lepton-number
($B$ and $L$, respectively). In an effective field theory these
operators are necessarily present, and can be dangerous even when
suppressed by the Planck scale, $M_\mathrm{Pl}$
[\refcite{Murayama:1994tc}].

It is also often sloppily said that $R$-parity prohibits proton decay
in SUSY theories. Though $R$-parity prohibits the
\emph{renormalizable} $B$- and $L$-violating operators, it still
allows the \emph{nonrenormalizable} superpotential terms
$\frac{1}{M}QQQL$ and $\frac{1}{M}\bbar{U}\bbar{U}\bbar{D}\bbar{E}$,
which contain dimension five operators that can lead to rapid proton
decay. In fact, with generic \orderone coefficients, weak scale squark
masses, and $M\sim M_\mathrm{Pl}$, the proton lifetime comes out to be
sixteen orders of magnitude below the current experimental limit! This
embarrassment has been called SUSY's ``dirty little secret''.

This ``dirty secret'' is most likely cleaned up by the
physics that generates flavor. If a broken flavor symmetry is the
source of the small Yukawa couplings, as in a Froggatt-Nielsen
model~[\refcite{Froggatt:1978nt}], then that same flavor symmetry will
govern the coefficients of the higher dimensional operators mentioned
above, allowing the suppression of their coefficients to be predicted.

In what follows we will survey the predictions for nucleon lifetimes
as computed in~[\refcite{Harnik:2004yp}] for the class of specific,
string-motivated models introduced in
[\refcite{Dreiner:2003yr}]. These models are based on a single,
anomalous $U(1)_X$ Froggatt-Nielsen flavor symmetry but \emph{do not}
require grand unification.

\section{Proton Decay Operators}

In GUT theories the exchange of
$X$ gauge bosons generates $B$- and $L$-violating four-fermion
operators suppressed by two powers of $M_\mathrm{GUT}$, yielding the
proton decay rate $\Gamma \sim
\frac{\alpha_\mathrm{GUT}^2}{M_\mathrm{GUT}^4}m_p^5$. For the standard
model, $M_\mathrm{GUT}\sim 10^{15}$ GeV leading to a proton lifetime
around $10^{31}$ years, well below the current limits which now exceed
$10^{33}$ years for many decay modes~[\refcite{Kearns}]. In a SUSY-GUT
the unification scale is a factor of 10 higher, suppressing the rate
from these dimension six operators by four more orders of magnitude,
evading the experimental constraint. However, colored Higgs exchange
generates dimension five couplings between fermions and their
superpartners which lead to four-fermion operators that are suppressed
by one power of $M_\mathrm{GUT}$ and one power of the scalar soft
mass, $m_\mathrm{soft}$. The proton decay rate becomes $\Gamma \sim
\frac{\alpha_\mathrm{GUT}^2}{M_\mathrm{GUT}^2 m_\mathrm{soft}^2}
m_p^5$. Since we expect $m_\mathrm{soft} \ll M_\mathrm{GUT}$, proton
decay from these operators is relatively enhanced and very dangerous,
excluding the minimal $SU(5)$ SUSY-GUT~[\refcite{Murayama:2001ur}].

Even without grand unification an effective field theory should
contain all allowed higher dimensional operators suppressed by
$M_\mathrm{Pl}$, including the dimension-5 $B$ and $L$ violating
operators mentioned above. They lead to proton decay with a rate
$\Gamma \sim \frac{\alpha^2}{M_\mathrm{Pl}^2 m_\mathrm{soft}^2}
m_p^5$. If such operators were present with \orderone coefficients it
would be disastrous. Therefore we need to consider the degree to which
these coefficients are suppressed by flavor physics.

\section{Flavor Model Framework}

In the class of models presented in~[\refcite{Dreiner:2003yr}] the
MSSM superfields are charged under a horizontal $U(1)_X$ symmetry that
is spontaneously broken when a flavon field, $A$, gets a nonzero VEV
generated by string dynamics.  Both the MSSM Yukawa terms and
the higher dimensional operators are then suppressed by the ratio
$\epsilon = \langle A \rangle /M_\mathrm{Pl}$ raised to the the
appropriate power necessary to conserve $U(1)_X$. The string dynamics
predicts $\epsilon \sim \sin\theta_C$.  The $X$-charges for the MSSM
superfields are restricted by sum rules that ensure anomaly
cancellation through the Green-Schwarz mechanism~[\refcite{Green:sg}],
and further constrained by requiring that they lead to the observed
fermion mass spectrum and mixings, including neutrinos and the MNS
matrix, and by requiring that $R$-parity be an exact, accidental
symmetry of the low energy theory. It is non-trivial that these
requirements can be simultaneously fulfilled. There are 24 distinct
models with these properties, parametrized by three integers $x$, $y$,
and $z$ that are related to $\tan\beta$, the CKM texture, and the MNS
texture, respectively. (See~[\refcite{Dreiner:2003yr,Harnik:2004yp}]
for details.)

\section{Results}
\label{sec:results}

The 24 models each make predictions for the parametric size
of the coefficients appearing in front of the dimension-five $B$ and
$L$ violating operators, allowing us to compute the lifetime of the
proton predicted by each model.

There are two types of uncertainties that enter into our
predictions. The first type of uncertainty comes from our ignorance of
$\beta_p$, the overall scale of the matrix elements for proton decay
as computed using the chiral Lagrangian
technique~[\refcite{Claudson:1981gh}], and our ignorance of the
superpartner mass scale $m_\mathrm{soft}$. These two uncertainties
will hopefully be reduced with time. The second type of uncertainty is
inherent in our effective field theory framework and comes from the
unknown \orderone coefficients that appear in front of each higher
dimensional operator. We estimate the effect of the unknown phases of
these coefficients by adding contributing amplitudes either
incoherently, destructively, or constructively.

\begin{figure}[ht]
  \centering \includegraphics[width=.7\columnwidth]{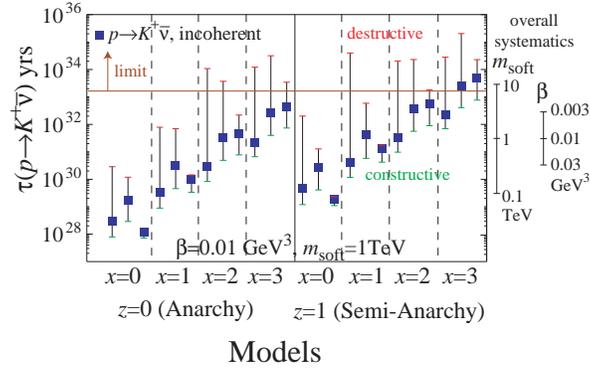}
  \caption{Plot of proton partial lifetime in years for the mode
  $p\rightarrow K^+\bar{\nu}$. Within each half $\tan\beta$ decreases
  from left to right.  The error bars are \emph{not} $1\sigma$ bars,
  but show the shift from incoherent addition of amplitudes (central
  value) due to destructive and constructive interference. The
  horizontal line shows the experimental lower limit of $1.6\times
  10^{33}$ years. The scales on the right show the overall shift
  caused by changing either $m_\mathrm{soft}$ or $\beta_p$ away from
  $m_\mathrm{soft}=1$ and $\beta_p=0.01$.}
  \label{fig:Knu-uncertainties}
\end{figure}

Figure~\ref{fig:Knu-uncertainties} shows the partial lifetimes for the
most constraining mode, $p\rightarrow K^+ \bar{\nu}$, for all 24
models labeled by the parameters $x$, $y$, and $z$. Already many of
the models are disfavored, unless they have significant cancellations
between contributing amplitudes. The models that are least constrained
are those with lower $\tan\beta$ (higher $x$). However, the
uncertainties in $\beta_p$ and $m_\mathrm{soft}$ can potentially
change the overall scale of the prediction by the factors shown
graphically to the right of Figure \ref{fig:Knu-uncertainties}.

\begin{figure}[tc]
  \centering
  \includegraphics[width=.65\columnwidth]{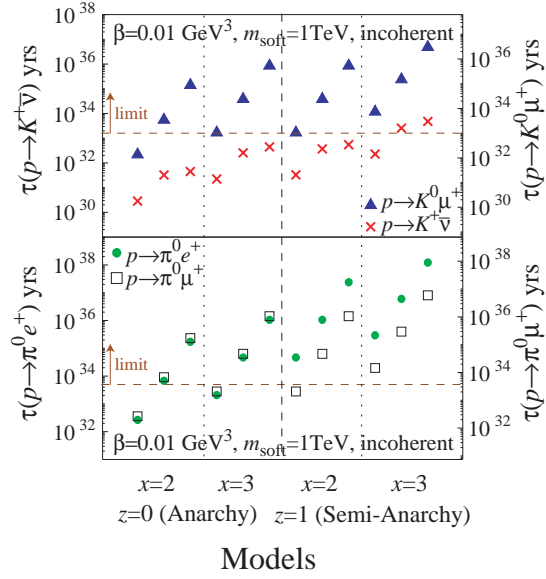}
  \caption{Comparison of proton lifetime in years for four different
    decay modes. The upper plot shows the computed lifetime for $p\to
    K^+ \bar{\nu}$ ($\times$, left axis) and $p\to K^0
    \mu^+$ ($\blacktriangle$, right axis). The lower
    plot shows $p\to \pi^0 e^+$ ($\bullet$, left axis)
    and $p\to \pi^0 \mu^+$ ($\square$, right axis).}
\label{fig:4modes}
\end{figure}

For the proton the next modes to appear after $p\to K^+ \bar{\nu}$ are
generally $p\to \pi^0 e^+$, $p\to \pi^0 \mu^+$, and $p\to K^0 \mu^+$.
In Figure~\ref{fig:4modes} we show the expected lifetime for those
four modes in the 12 models with $\tan\beta \lesssim 10$. We see that
most models which survive the constraint from $p\to K^+ \bar{\nu}$
have a lifetime for $p\to \pi^0\mu^+$ that is within two or three
orders of magnitude of the experimental bound, while $p\to K^0 \mu^+$
is only slightly larger, and $p\to \pi^0 e^+$ can potentially be
smaller. This raises the exciting possibility of two or three decay
modes being detected in the coming round of experiments.

\begin{figure}[htc]
\centering \includegraphics[width=.7\columnwidth]{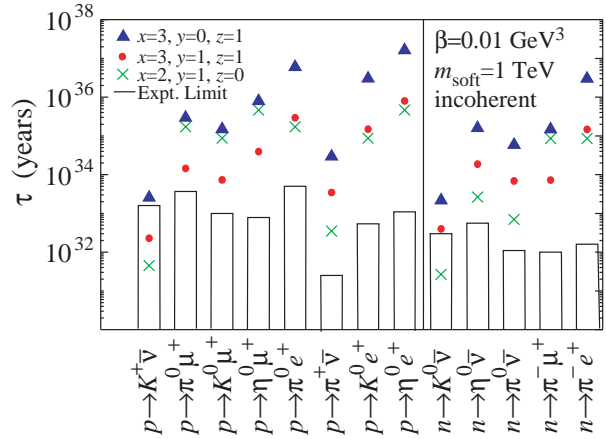}
\caption{Plot of nucleon lifetime in years for eight proton decay
modes (left side) and five neutron decay modes (right side).  The
different symbols represent different $U(1)_X$ charge assignments. The
experimental limit for each mode is shown as a vertical column.}
\label{fig:3models}
\end{figure}

Figure~\ref{fig:3models} shows the partial lifetimes in 11 proton and
neutron decay modes for three models, illustrating how various modes
can discriminate between models. For example, any mode involving a
muon in the final state can differentiate Model 1 from Models 2 and 3.

\section{Conclusion}
\label{sec:conclusion}

Focusing on a class of string motivated Froggatt-Nielsen
models that explain the masses and mixings of all SM fermions while
automatically enforcing $R$-parity, we have shown that nucleon decay
is a powerful probe of Planck scale physics. In these models Planck
suppressed operators lead to nucleon lifetimes that are
generically right near the current experimental limits, even without
grand unification. Since current bounds constrain many of the 24
models of this type, we conclude that proton decay is already
probing physics at the Planck scale.

\section*{Acknowledgments}
DTL thanks R.~Harnik, H.~Murayama, and M.~Thormeier for
fruitful collaboration. This work was supported in part by the DOE
under contract DE-AC03-76SF00098 and in part by NSF grant
PHY-0098840.


%
%
%
%


\begin{thebibliography}{0}

\bibitem{Murayama:1994tc}
H.~Murayama and D.~B.~Kaplan,
Phys.\ Lett.\ B {\bf 336}, 221 (1994)
[arXiv:hep-ph/9406423].

\bibitem{Froggatt:1978nt}
C.~D.~Froggatt and H.~B.~Nielsen,
Nucl.\ Phys.\ B {\bf 147}, 277 (1979).


\bibitem{Harnik:2004yp}
R.~Harnik, D.~T.~Larson, H.~Murayama and M.~Thormeier,
arXiv:hep-ph/0404260.

\bibitem{Dreiner:2003yr}
H.~K.~Dreiner, H.~Murayama and M.~Thormeier,
arXiv:hep-ph/0312012.

\bibitem{Kearns}
E.~Kearns, Talk at Snowmass 2001,\\
\mbox{http://hep.bu.edu/\~{}kearns/pub/kearns-pdk-snowmass.pdf}.

\bibitem{Murayama:2001ur}
H.~Murayama and A.~Pierce,
Phys.\ Rev.\ D {\bf 65}, 055009 (2002)
[arXiv:hep-ph/0108104].

\bibitem{Green:sg}
M.~B.~Green and J.~H.~Schwarz,
Phys.\ Lett.\ B {\bf 149}, 117 (1984).

\bibitem{Claudson:1981gh}
M.~Claudson, M.~B.~Wise and L.~J.~Hall,
Nucl.\ Phys.\ B {\bf 195}, 297 (1982).

\end{thebibliography}
\end{document}